\documentclass[aps,floatfix,twocolumn,showpacs,preprintnumbers]{revtex4}

\usepackage{mathrsfs}
\usepackage{amsmath, amssymb, amsfonts, bbm}
\usepackage{graphics,epsfig,color,subfigure}

\newcommand{\be}{\begin{equation}}
\newcommand{\ee}{\end{equation}}
\def\bea{\begin{align}}
\def\ena{\end{align}}

\def\beqa{\begin{eqnarray}}
\def\enqa{\end{eqnarray}}

\begin{document}

\title{All you need is N:  Baryon spectroscopy in two large N limits}

\author{Aleksey Cherman}
\email{alekseyc@physics.umd.edu}
\affiliation{Center for Fundamental Physics, Department of
Physics, University of Maryland, College Park, MD 20742-4111}

\author{Thomas D. Cohen}
\email{cohen@physics.umd.edu}
\affiliation{Maryland Center for Fundamental Physics, Department of
Physics, University of Maryland, College Park, MD 20742-4111}

\author{Richard F.  Lebed}
\email{Richard.Lebed@asu.edu}
\affiliation{Department of Physics, Arizona State University, Tempe,
AZ 85287-1504}

\date{June 2009}

\begin{abstract}
The generalization of QCD to many colors is not unique; each distinct
choice corresponds to a distinct $1/N_c$ expansion.  The familiar
't~Hooft $N_c \! \to \infty$ limit places quarks in the fundamental
representation of SU($N_c$), while an alternative approach places
quarks in its two-index antisymmetric representation.  At $N_c=3$
these two $1/N_c$ expansions coincide.  We compare their predictions
for certain observables in baryon spectroscopy, particularly mass
combinations organized according to SU(3) flavor breaking.  Each large
$N_c$ limit generates an emergent spin-flavor symmetry that leads to
the vanishing of particular linear combinations of baryon masses at
specific orders in the expansions.  Experimental evidence shows that
these relations hold at the expected orders regardless of which large
$N_c$ limit one uses, suggesting the validity of either limit in the
study of baryons.  We also consider a hybrid large $N_c$ limit in
which one flavor is taken to transform in the two-index antisymmetric
representation and the rest of the flavors are in the fundamental
representation.  While this hybrid large $N_c$ limit is theoretically
attractive, we show that for a wide class of observables it faces some phenomenological difficulties.
\end{abstract}

\preprint{DOE-40762-455}
\preprint{INT-PUB-09-027}

%

\maketitle

\section{Introduction}

The study of large $N_c$ QCD and the $1/N_c$ expansion is nearly as
old as QCD itself~\cite{'tHooft:1973jz}.  While much of its
significance lies in the formal theoretical domain, the $1/N_c$
expansion has proven to be an important source of qualitative and
semi-quantitative insights into hadronic physics.  Underlying the
approach is the premise that a world with a very large number of
colors is not too different from a world with $N_c=3$.  Thus it is
sensible to compute some hadronic properties of interest in the $N_c
\rightarrow \infty$ world and then to improve predictions
systematically by taking $1/N_c$ corrections into account.  Implicit
in this formulation is the assumption of a unique and well-defined
large $N_c$ world.  However, the extrapolation from $N_c=3$ to $N_c
\rightarrow \infty$ has long been known not to be
unique~\cite{Corrigan:1979xf}; one can construct distinct theories
that agree at $N_c=3$ but differ at $N_c=\infty$.  Each distinct
extrapolation corresponds to a distinct $1/N_c$ expansion for physical
quantities.  It is thus important to uncover {\it which\/} large $N_c$
limits are appropriate for various quantities of phenomenological
interest.

The large $N_c$ limit is usually taken by placing quarks in the
fundamental representation F of SU($N_c$), as occurs for the physical
case $N_c = 3$.  But $N_c = 3$ is exceptional because the
(anti)fundamental representation and two-index antisymmetric
representation AS [$N_c(N_c - 1)/2$ dimensional in general] are
isomorphic: we can associate each AS quark $q^{j k}$ with an F quark
$q_i$ via $q_i = \frac 1 2 \epsilon_{ijk} q^{j k}$.  For example, an
$N_c = 3$ red-blue quark in the AS representation is equivalent to an
anti-green quark in $\overline{\rm F}$.  Exchanging quarks with
antiquarks, one sees that this ``QCD$_{\rm AS}$'' at $N_c=3$ is just
ordinary ``QCD$_{\rm F}$".  For $N_c > 3$ however, QCD$_{\rm AS}$ is
clearly different from QCD$_{\rm F}$.  Each of them has a distinct
large $N_c$ limit, which we refer to respectively as the large
$N_c^{\mathrm{AS}}$ limit and the large $N_c^{\mathrm{F}}$ limit.
When generically referring to either large limit, we continue to use
$N_c$ without a superscript.

Since these two limits describe different large $N_c$ worlds, our
$N_c=3$ world may in effect be described more accurately by one of
them, in the sense that one of the $1/N_c$ expansions may turn out to
be more useful for describing some sets of observables.  Indeed, it is
not {\it a priori\/} clear if either limit is viable for a given set
of observables.

Compelling theoretical reasons underlie the study of AS Dirac quarks
and the large $N_c^{\mathrm{AS}}$ limit.   A key aspect is the large
$N_c$ connection of QCD$_{\rm AS}$ to QCD$_{\rm Adj}$, which is
Yang-Mills theory with Majorana quarks in the adjoint (Adj)
representation.  Armoni, Shifman, and Veneziano~\cite{Armoni:2003gp}
have found an orientifold equivalence between the respective large
$N_c^{\mathrm{AS}}$ and $N_c^{\mathrm{Adj}}$ limits.  More precisely,
for a certain set of observables, the differences between QCD$_{\rm
AS}$ and QCD$_{\rm Adj}$ as $N_c \to \infty$ are subleading in
$1/N_c$, a fact that is interesting in its own right.  Moreover,
QCD$_{\rm Adj}$ with a single massless flavor is supersymmetric,
allowing one to use SUSY techniques to compute some observables
explicitly in one-flavor QCD$_{\rm Adj}$, and equate them at large
$N_c$ to corresponding ones in one-flavor QCD$_{\rm
AS}$~\cite{Armoni:2003gp}.

While the large $N_c^{\mathrm{AS}}$ limit is clearly theoretically
interesting, determining its phenomenological relevance in describing
the real $N_c=3$ world is equally important.  This paper compares the
phenomenological viability of the large $N_c^\mathrm{AS}$ limit to
that of the standard large $N_c^{\textrm{F}}$ limit.  We focus on
testing the two large $N_c$ limits with observables from baryon
spectroscopy, which can be used to systematically probe the
differences between the two limits.  Such a study is possible because
a spin-flavor symmetry emerges in the baryon sector at large
$N_c$~\cite{emergent,DM,Dashen:1993jt}, and when combined with the
hierarchy associated with SU(3)$_{\rm flavor}$ breaking, this symmetry
makes quantitative predictions for particular combinations of baryon
masses.

Each of these baryon mass relations has been previously shown to hold
to the expected accuracy for the usual large $N_c^{\mathrm{F}}$ limit
of QCD~\cite{Jenkins:1995td}.  In particular,
Ref.~\cite{Jenkins:1995td} showed that these relations hold far better
than if one had merely used SU(3)$_{\rm flavor}$ relations.  The large
$N_c^{\mathrm{F}}$ world is therefore sufficiently close to the
$N_c=3$ world that the $1/N_c$ expansion has predictive power for the
mass observables.

In this paper, we use the large $N_c$ spin-flavor symmetry together
with SU(3)$_{\rm flavor}$ breaking to show that the predictions of the
large $N_c^{\mathrm{AS}}$ expansion match the experimental data for
the mass observables much better than those using SU(3)$_{\rm flavor}$
breaking only.  Furthermore, we show that the large
$N_c^{\mathrm{AS}}$ expansion explains the patterns in the data
approximately as well as the large $N_c^{\mathrm{F}}$ expansion.  Thus
one sees that, while the data require \emph{some\/} large $N_c$ limit,
it does not determine \emph{which\/} large $N_c$ expansion is the
correct one.  The large $N_c^{\mathrm{F}}$ world and the large
$N_c^{\mathrm{AS}}$ world are about equally close to the $N_c=3$
world, at least as determined by the baryon mass spectrum.

As defined above, the large $N_c^{\mathrm{AS}}$ and $N_c^{\mathrm{F}}$
limits of QCD both treat all quark flavors symmetrically, and we focus
on these two flavor-symmetric large $N_c$ limits for most of this
paper.  We also briefly discuss a hybrid large $N_c$ limit in which
one flavor is singled out to transform in the AS representation, while
the rest remain in the F representation.  This hybrid
``Corrigan-Ramond'' expansion has a number of theoretical attractions,
but as shown in Sec.~\ref{sec:CRlargeN}, the lack of flavor symmetry
for $N_c \neq 3$ makes its phenomenological utility questionable for a
large class of observables that includes (for instance) baryon mass
spectra.

The organization of the paper is as follows.  In
Sec.~\ref{sec:BaryonGeneralities}, we discuss some general
phenomenological features of the large $N_c^{\mathrm{AS}}$ limit and
define the phenomenologically appropriate large $N_c$ generalization
of baryons in QCD$_{\mathrm{AS}}$.  Next, in
Sec.~\ref{sec:SpinFlavor}, we briefly review the emergent spin-flavor
symmetry for baryons in the large $N_c$ limit, and discuss its
application to the $N_c^{\mathrm{AS}}$ limit, and in
Sec.~\ref{sec:Data}, we confront the predictions of the two large
$N_c$ limits with experimental data.  We then discuss the
phenomenology of a hybrid large $N_c$ limit in
Sec.~\ref{sec:CRlargeN}, and summarize in Sec.~\ref{sec:Conclusions}.

\section{Baryons in the large $N_c^{\mathrm{AS}}$ limit}
\label{sec:BaryonGeneralities}

In many ways the $1/N_c^{\mathrm{F}}$ and $1/N_c^{\mathrm{AS}}$
expansions are similar: Both large $N_c$ limits are planar, contain an
infinite number of narrow mesons, and so on.  However, internal F
quark loops are suppressed in the large $N_c^{\mathrm{F}}$ limit but
not in the large $N_c^{\mathrm{AS}}$ limit.  This feature results from
gluons and AS quarks sharing one important property: Summing over
their internal quantum numbers means including $O(N_c^2)$ color index
values, whereas sums over F quarks include only $N_c$ index values.

Thus the OZI rule and all of its phenomenological implications ({\it
e.g.}, a small $\rho$-$\omega$ mass splitting) naturally emerge in the
$N_c^{\mathrm{F}}$ limit, but \emph{not\/} in the $N_c^{\mathrm{AS}}$
limit.  However, this particular phenomenological virtue of the
$N_c^{\mathrm{F}}$ limit induces a corresponding vice, as indicated by
cases like the $\pi$-$\eta^\prime$ mass splitting in which Zweig's
rule fails badly.  Which of the two expansions is preferable depends
on the observable.
 
In this paper we focus on mass observables associated with baryon
spectroscopy.  We now define what we mean by ``baryons'' in the two
large $N_c$ limits, and discuss some of their properties at large
$N_c$.  In the large $N_c^{\mathrm{F}}$ limit, the construction of
color-singlet baryons $B_{\mathrm{F}}$ with $N_c$ quarks is
straightforward~\cite{Witten:1979kh}:
\be \label{Witten}
B_{\mathrm{F}} \sim \epsilon^{i_1 , \, i_2, \, \ldots \, , \, i_{N_c}}
q_{i_1} q_{i_2} \! \cdots q_{i_{N_c}} \, .
\ee
The situation is somewhat more complicated in the large
$N_c^{\mathrm{AS}}$ limit, since more than one way exists to construct
large $N_c^{\mathrm{AS}}$ color-singlet
``baryons''~\cite{Bolognesi:2006ws}.  One may couple $N_c$ quarks with
color indices contracted as follows:
\begin{equation}
B_{\phi} \sim  \epsilon^{j_1, \, j_2, \, \cdots, \, j_{N_c}}
q_{j_1, \, j_2} q_{j_3, \, j_4} \cdots q_{j_{N_{c-1}}, \, j_{N_c}} \,
,
\label{phi}
\end{equation}
which represents a sum of terms, each one containing $N_c/2$ quark
fields.  A natural alternative construction uses sums involving two epsilon tensors and $N_c(N_c-1)/2$ quarks, and is fully antisymmetric under the interchange of any two quarks\cite{Bolognesi:2006ws}.   For $N_c=3$, this can be written simply as 
\begin{equation}
B_{\psi} \sim (\epsilon_{i_2,j_2,i_1}\epsilon_{i_3,j_3,j_1}-\epsilon_{i_3,j_3,i_1}\epsilon_{i_2,j_2,j_1})q^{i_1 ,j_1}q^{i_2,j_2}q^{i_3 ,j_3},
\label{psi}
\end{equation}
and one must keep in mind that the indices of the two-index quarks are to be antisymmetrized.  The methods of Ref.~\cite{Bolognesi:2006ws} can be used to straightforwardly write similar expressions for general $N_c$, but the resulting expressions are somewhat cumbersome and we do not show them here.

Each of these combinations of quarks can reasonably be described as a
baryon since each carries nonzero baryon number.  At the same time,
these two classes of baryon have dramatically different properties.
For example, denoting by $\mathcal{B}_{\phi}$ and $\mathcal{B}_{\psi}$
the baryon number of $B_{\phi}$ and $B_{\psi}$ as determined by quark
number, one has
\begin{equation}
\frac{\mathcal{B}_{\phi}}{\mathcal{B}_{\psi}} = \frac{1}{N_c-1} \, .
\end{equation}
It might appear that an ambiguity has arisen as to which of these
types of baryons should be taken as the large $N_c^{\mathrm{AS}}$
analog of our usual $N_c=3$ baryons.  In fact, this is not the case:
By construction, $B_{\phi}$ baryons only exist for $N_c$ even (and
have other questionable physical properties~\cite{Bolognesi:2006ws}).
Thus the $B_{\psi}$ baryon is the better generalization of the $N_c=3$
case.  Accordingly, we take the baryon number per quark to be $[N_c
(N_c -1)/2]^{-1}$.  Extrapolating solely to odd values of large $N_c$
eliminates $B_{\phi}$ baryons from further consideration.  When
discussing QCD$_{\mathrm{AS}}$ ``baryons'' in the remainder of this
paper, we mean $B_{\psi}$ baryons.

Baryons at large $N_c^{\mathrm{AS}}$ behave as one might expect: As
the number of quarks in a baryon scales like $N_c^2$, the baryon mass
also scales like $N_c^2$~\cite{Bolognesi:2006ws}.  However, showing
that the interactions between AS quarks are consistent with this
counting involves somewhat more subtle combinatorics than for F
quarks~\cite{Cherman:2006iy}.  Witten noted long
ago~\cite{Witten:1983tx} that baryon properties in the conventional
large $N_c^{\mathrm{F}}$ limit scale as in a generic soliton model
with the mesonic coupling constant scaling as $g \sim 1/N_c^{1/2}$.
Similarly, one expects that baryon properties in the large
$N_c^{\mathrm{AS}}$ limit scale as in a generic soliton model with the
mesonic coupling constant scaling as $g \sim 1/N_c$.  It has been
shown that observables in QCD$_{\rm AS}$ Skyrme-type models
self-consistently obey the correct $1/N_c$
scaling~\cite{Cherman:2006iy}.

\section{Spin-flavor symmetry as a probe}
\label{sec:SpinFlavor}

The central question of this paper is the extent to which the large
$N_c^{\mathrm{AS}}$ limit is viable for phenomenology, specifically
for baryon spectroscopy.  Qualitatively, the large $N_c^{\mathrm{AS}}$
limit differs from the large $N_c^{\mathrm{F}}$ limit in three ways.
First, the generic scaling of many quantities associated with meson
and baryon observables is different, and many quantities behave as if
the expansion parameter in the meson and baryon sector is $1/N_c^2$
rather than $1/N_c$.  For instance, meson decay constants scale as
$f_m^2 \sim N_c^2$ rather than $f_m^2 \sim N_c$, with analogous
changes for many other quantities~\cite{Cherman:2006iy}.  Second, the
detailed dynamics are different in the two theories: One cannot simply
compute quantities in QCD$_{\mathrm{F}}$ at large $N_c^{\mathrm{F}}$
and replace $N_c$ by $N_c(N_c-1)/2$ to obtain the large
$N_c^{\mathrm{AS}}$ limit of QCD$_{\mathrm{AS}}$.  At a diagrammatic
level, this distinction is reflected by rather different
combinatorics.  It is largely for this reason that we take the
expansion parameter to be $1/N_c^2$ and not $[N_c(N_c-1)/2]^{-1}$.
Third, observables associated with internal quark loops, {\it e.g.},
strangeness form factors of the proton~\cite{Cherman:2006xa}, are
suppressed in the large $N_c^{\mathrm{F}}$ limit but not in the large
$N_c^{\mathrm{AS}}$ limit.

How to study all three of these distinctions systematically is not
obvious.  For example, identifying loop effects in baryons is not
generally straightforward, with the exception of using the strangeness
content of the nucleon as a probe.  Unfortunately, rather limited data
of this sort exist, making the identification of patterns difficult.
Evidence suggests that strange-quark matrix elements are typically
quite small (for some recent reviews, see
Refs.~\cite{Beise:2004py,Beck:2001yx,Beck:2001dz}), in contrast to
expectations from the large $N_c^{\mathrm{AS}}$ limit.  However, it
has been argued that the strange scalar matrix elements are
large~\cite{Cherman:2006xa}, as expected in the large
$N_c^{\mathrm{AS}}$ but not the large $N_c^{\mathrm{F}}$ limit.  Given
these results, real conclusions remain elusive.  Distinguishing
between the limits based upon detailed dynamical differences is even
more difficult; short of dedicated lattice studies of baryons built
around either limit, no obvious scheme suggests itself.  We therefore
focus on the first difference, the distinct $1/N_c$ scaling behaviors
of certain observables in the two expansions.

Fortunately, the scaling is easily probed by considering the
spectroscopy of low-lying baryons: the octet and decuplet.  The
existence of an emergent spin-flavor symmetry SU($2N_f$) at large
$N_c$~\cite{emergent,DM,Dashen:1993jt}, which places all these baryons
in a single multiplet (the old-fashioned {\bf 56} when $N_c = 3$ and
$N_f = 3$), provides a tool for comparing the two large $N_c$ limits.
This symmetry, which implies mass relations that hold to various
orders in a $1/N_c$ expansion, can be uncovered by studying
pion-nucleon scattering.  The Born and cross-Born graphs in that
process are each proportional to $(g_A/f_\pi)^2$, which is large in
either large $N_c$ limit, scaling as either $N_c$ or $N_c^2$ in
QCD$_{\mathrm{F}}$ and QCD$_{\mathrm{AS}}$, respectively.  In either
case, these graphs individually would violate unitarity at large
$N_c$, which necessitates cancellations between them in order to
maintain a consistent theory.  The consistency conditions implicit in
these cancellations yield the spin-flavor symmetry at large $N_c$.

At finite $N_c$ the cancellations are only partial, but a remarkable
feature survives: If the underlying $N_c \to \infty$ symmetry is
exact, the cancellations hold at next-to-leading order in the $1/N_c$
expansion~\cite{DM,Dashen:1993jt}.  For the standard large
$N_c^{\mathrm{F}}$ limit this means that the corrections first appear
at order $1/N_c^2$; for the large $N_c^{\mathrm{AS}}$ limit, the order
is $1/N_c^4$.  Moreover, one can use explicit breaking of
SU(3)$_\mathrm{flavor}$ as a lever to probe the $1/N_c$ expansion.
When SU(3)$_\mathrm{flavor}$ is broken at first order (due to the
strange quark mass, say), then the spin-flavor symmetry is also
broken---but in a controlled way.  By studying both the
SU(3)$_\mathrm{flavor}$ and $1/N_c$ scalings of operators giving rise
to baryon masses, one can show that different linear combinations of
masses that vanish in the SU(3)$_{\mathrm{flavor}}$ limit can have
different $N_c$ scalings: Some are leading order (LO), some are
next-to-leading order (NLO), and some are next-to-next-to-leading
order (NNLO).  Since the phenomenological analysis sets $N_c = 3$ at
the end of the calculation, the series stops at NNLO\@.  Some details
of this operator analysis appear in the Appendix.

Using these methods, Ref.~\cite{Jenkins:1995td} studied the baryon
mass relations in the standard large $N_c^{\mathrm{F}}$ limit.  The
patterns seen in~\cite{Jenkins:1995td} are quite instructive, and give
some of the strongest evidence to date for the validity of the $1/N_c$
expansion in hadronic physics.  The expansion has real predictive
power: Relations dependent on SU(3)$_{\mathrm{flavor}}$ breaking that
hold at LO are not satisfied as well as those that hold at NLO, which
in turn are not satisfied as well as those at NNLO.  Even more
impressively, if one parameterizes these violations by constants times
appropriate powers of $1/N_c$ and inserts $N_c=3$, one finds that the
coefficients are of natural size.

What would one expect upon redoing this analysis using the large
$N_c^\mathrm{AS}$ limit?  It is not hard to see that the general
pattern determining which relations are well satisfied and which ones
are not is unaltered, because the emergent large $N_c$ spin-flavor
symmetry---a direct consequence of the nontrivial $N_c$ scaling of
meson-baryon coupling constants---is common to both large $N_c$
limits.

One might worry that the lack of suppression of quark loops in the
large $N_c^{\mathrm{AS}}$ limit could change the scaling of the
suppressed baryon mass combinations compared to the large
$N_c^{\mathrm{F}}$ limit.  To see that this does not happen, recall
that the emergent SU(2$N_f$) symmetry mandates the degeneracy of all
ground-state baryons in both large $N_c$ limits.  While quark loops
certainly contribute to baryon masses, they do not change the scaling
of baryon mass differences at leading order.  Incorporating explicit
SU(3)$_\mathrm{flavor}$ breaking does not change these conclusions,
because each possible operator has a leading-order contribution in
$1/N_c$ that does not include a quark loop.  Thus the spin-flavor
symmetry predicts that the same baryon mass combinations are
suppressed at large $N_c$ for both large $N_c$ limits.  However, the
\emph{degree\/} of $1/N_c$ suppressions is different for the
large $N_c^{\mathrm{AS}}$ and $N_c^{\mathrm{F}}$ limits.

Since the relevant expansion is in powers of $1/N_c^2 = 1/9$ for the
large $N_c^{\mathrm{AS}}$ limit versus $1/N_c = 1/3$ for large
$N_c^{\mathrm{F}}$, one might expect to find coefficients that are no
longer natural.  Were this true, one would conclude that the large
$N_c^{\mathrm{AS}}$ limit is less useful for describing baryons than
the large $N_c^{\mathrm{F}}$ limit.  Remarkably, as shown in the
remainder of this paper, the data show that \emph{both\/} expansions
give natural coefficients.  Such a surprising conclusion is possible
because spin-flavor symmetry for $N_c = 3$ gives access to only two
subleading orders in the expansion.  Based upon this evidence,
\emph{both\/} expansions appear to be viable.

\section{Confronting the two large $N_c$ limits with data}
\label{sec:Data}

The suppressed baryon mass combinations are classified according to
the isospin $I$ transformation properties of operators that contribute
to them.  References~\cite{Jenkins:1995td,Jenkins:2000mi} examine the
$I=0, 1, 2$ mass combinations that are suppressed in the combined
$1/N_c$ and flavor-breaking analysis.  Unfortunately, the experimental
errors in the isospin-breaking $I=1,2$ mass relations are too large to
allow a meaningful comparison that could discriminate between the two
large $N_c$ limits.  Thus in this paper, we focus on the $I=0$ mass
relations.

The specific $I=0$ mass combinations analyzed here are defined as
\begin{eqnarray}
N_0        & = & \frac 1 2 \left( p + n \right), \nonumber\\
\Sigma_0   & = & \frac 1 3 \left( \Sigma^+ + \Sigma^0 + \Sigma^-
\right), \nonumber\\
\Xi_0      & = & \frac 1 2 \left( \Xi^0 + \Xi^- \right), \nonumber\\
\Delta_0   & = & \frac 1 4 \left( \Delta^{++} +\Delta^+ +\Delta^0
+\Delta^- \right), \\
\Sigma^*_0 & = & \frac 1 3 \left( \Sigma^{*+} +\Sigma^{*0}
+\Sigma^{*-} \right), \nonumber\\
\Xi^*_0    & = & \frac 1 2 \left( \Xi^{*0} + \Xi^{*-} \right).
\nonumber
\end{eqnarray}
As shown in Ref.~\cite{Jenkins:1995td} (and briefly reviewed in the
Appendix), these mass combinations satisfy certain relations
summarized in Table~\ref{MassRelations}.  Each large $N_c$ limit
predicts that these relations are suppressed by specific powers of
$1/N_c$ and powers of the SU(3)$_{\rm flavor}$ symmetry-breaking
parameter $\epsilon$.

To compare the mass relations to the theoretical predictions from the
$1/N_c$ and SU(3)$_{\rm flavor}$-breaking analysis, one forms
dimensionless ratios $R_i$ from the mass combinations $M_i$.  Given an
$M_i$ from Table~\ref{MassRelations} ($i=1,2,\ldots7$), the ratios are
defined by $R_i \equiv M_i/(M_i^\prime /2)$, where $M_i^\prime$ is the
same combination of masses as in $M_i$, except that each numerical
coefficient is replaced by its absolute magnitude.  The ratios $R_i$
defined in this way measure the experimental size of each mass
combination $M_i$ relative to the weighted average $M_i^\prime$ of all
the masses appearing in $M_i$.

We then compute the expected theoretical suppression $S_i$ for each
$R_i$, which consists of the powers of $\epsilon$ and $1/N_c$
appropriate to $M_i$, as compiled in Table~\ref{MassRelations}.  Each
$M_i^\prime$, being a baryon average mass, is $O(N_c^1)$ in the large
$N_c^{\rm F}$ limit and $O(N_c^2)$ in the large $N_c^{\rm AS}$ limit.
If the combined SU(3)$_{\rm flavor}$-breaking and $1/N_c$ expansions
are effective, in the sense that a fit to data produces coefficients
of natural size, then one expects each $R_i/S_i$ to be of order unity.
A more telling metric for testing this hypothesis is the set of {\it
accuracy factors} $A_i \equiv \log(R_i/S_i)$, since this expression
encodes the fact that, as far as naturalness is concerned, deviations
from unity like $R_i / S_i \approx 2$ and $R_i / S_i \approx 1/2$ are
equally significant.  If the theoretical and experimental suppressions
agree, one expects that the $A_i$ should lie in a range of about $-1
\lesssim A_i \lesssim 1$, since $1 \approx \log 3$ (the amount that
would be caused by an inappropriate factor of $N_c$ in the large
$N_c^{\mathrm{F}}$ limit).

To compare the data with the theoretical predictions, one must also
estimate the size of the SU(3)$_{\rm flavor}$-breaking parameter
$\epsilon$.  To obtain an estimate independent of large $N_c$ effects
and sensitive only to SU(3)$_{\rm flavor}$-breaking effects, we take
$\epsilon$ to be the average of the mass splittings between $N_0$ and
$\Lambda, \, \Sigma_0$, and $\Xi_0$.  Specifically, we define
\be
\epsilon \equiv \frac{1}{3} \sum_{i=1}^{3}
{\frac{B_i-N_0}{(B_i+N_0)/2}} \;,
\ee
where $B_i = \Lambda, \, \Sigma_0, \, \Xi_0$.  This particular choice,
among many other possible options, gives $\epsilon \approx 0.25$,
which is the same value used in Ref.~\cite{Jenkins:1995td}.

\begin{widetext}
\begin{center}
\begin{table}[ht]
\begin{tabular}{| c | c | c | c | }
\hline
      & Mass Combination & Large $N_c^{\mathrm{F}}$ suppression &
      Large $N_c^{\mathrm{AS}}$ suppression \\
\hline
$M_1$ &  $5(2N_0+ 3\Sigma_0 +\Lambda +2\Xi_0) -4(4\Delta_0
+3\Sigma^*_0 +2\Xi^*_0 +\Omega)$  & $1/N_c$ & $1/N_c^2$  \\
\hline
$M_2$ & $5(6N_0 -3\Sigma_0 +\Lambda -4\Xi_0) -2(2\Delta_0 -\Xi^*_0
-\Omega) $ & $\epsilon$ & $\epsilon$ \\
\hline
$M_3$ & $N_0 -3\Sigma_0 +\Lambda +\Xi_0$ & $\epsilon / N_c$ &
$\epsilon / N_c^2$ \\
\hline
$M_4$ & $ (-2N_0 -9\Sigma_0+3\Lambda + 8\Xi_0) +2(2\Delta_0 -\Xi^*_0
-\Omega)$ & $\epsilon / N_c^2$ & $\epsilon / N_c^4$ \\
\hline
$M_5$ & $35(2N_0 -\Sigma_0 -3\Lambda +2\Xi_0) -4(4\Delta_0
-5\Sigma^*_0 -2\Xi^*_0 +3\Omega)$ & $\epsilon^2 / N_c$ & $\epsilon^2 /
N_c^2$ \\
\hline
$M_6$ & $ 7 (2N_0 -\Sigma_0-3\Lambda + 2\Xi_0) -2(4\Delta_0
-5\Sigma^*_0 -2\Xi^*_0 +3\Omega) $ & $\epsilon^2 / N_c^2$ &
$\epsilon^2 / N_c^4$ \\
\hline
$M_7$ & $\Delta_0 - 3 \Sigma^*_0 + 3 \Xi^*_0 - \Omega$ & $\epsilon^3 /
N_c^2$ & $\epsilon^3 / N_c^4$ \\
\hline
\end{tabular}
\caption{The $I=0$ mass relations and their theoretical suppression
factors.}
\label{MassRelations}
\end{table}
\end{center}
\end{widetext}

\begin{figure}[ht]
\centering
\includegraphics[scale=1]{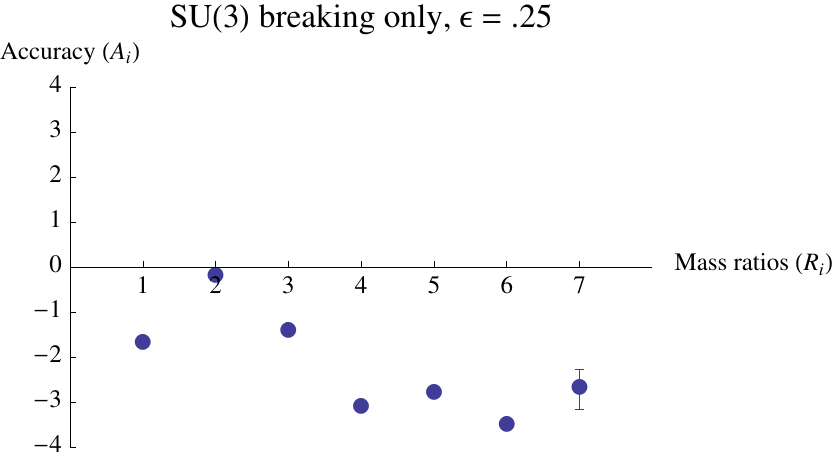}
\caption{Plot of the accuracy factors $A_i$ for $I=0$ mass ratios
$R_i$ when only SU(3)$_{\rm flavor}$-breaking suppressions are
included.}
\label{FlavorOnlyPlot}
\end{figure}
\begin{figure}[ht]
\centering
\includegraphics[scale=1]{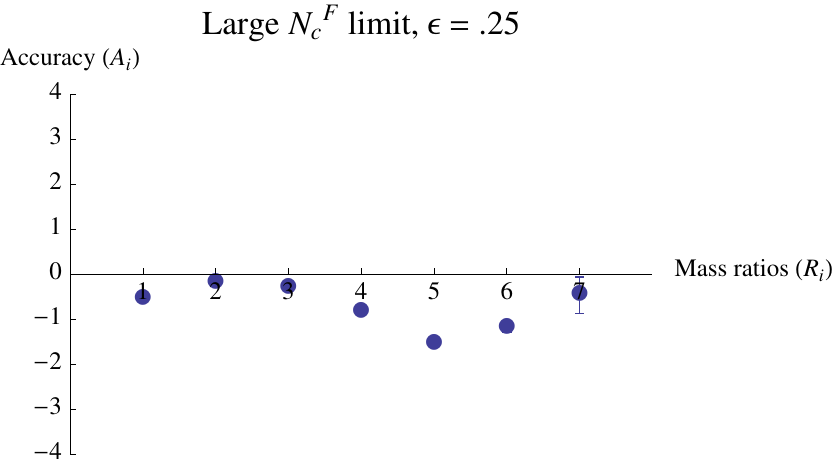}
\caption{Plot of the accuracy factors $A_i$ for $I=0$ mass ratios
$R_i$ when the combined SU(3)$_{\rm flavor}$-breaking and large
$N_c^{\mathrm{F}}$ limit suppression factors are used.}
\label{tHooftPlot}
\end{figure}
\begin{figure}[ht]
\centering
\includegraphics[scale=1]{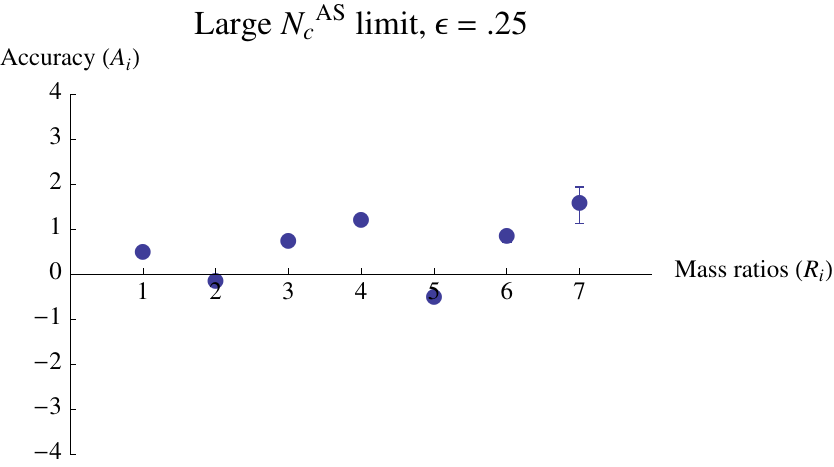}
\caption{Plot of the accuracy factors $A_i$ for $I=0$ mass ratios
$R_i$ when the combined SU(3)$_{\rm flavor}$-breaking and large
$N_c^{\mathrm{AS}}$ limit suppression factors are used. }
\label{ASPlot}
\end{figure}

In Fig.~\ref{FlavorOnlyPlot} we plot the accuracy factors $A_i$ for $i
= 1,2,\ldots,7$ when one includes only SU(3)$_{\rm flavor}$ breaking
in the theoretical suppression factors $S_i$.  In
Fig.~\ref{tHooftPlot} we plot the $A_i$ using theoretical suppression
factors $S_i$ obtained from the large $N_c^{\mathrm{F}}$ limit, while
in Fig.~\ref{ASPlot} we plot the $A_i$ using $S_i$ obtained from the
large $N_c^{\mathrm{AS}}$ limit.  In all three plots, the error bars
are due to experimental uncertainties in the baryon mass combinations
$M_i$ (and only visible for $M_7$).  

Figure~\ref{FlavorOnlyPlot} clearly shows that the accuracy factors
$A_i$ do \emph{not\/} assume natural values when one uses only
SU(3)$_{\mathrm{flavor}}$-breaking effects for the suppression factors
$S_i$.  It is also clear from Figs.~\ref{tHooftPlot}--\ref{ASPlot}
that the $A_i$ assume natural values for \emph{both\/} large $N_c$
limits.  Different definitions for the value of $\epsilon$ do not
change this result.

Thus, from the plots in Figs.~\ref{FlavorOnlyPlot}--\ref{ASPlot}, one
sees that a large $N_c$ limit explains the global pattern of
suppressions in the baryon mass splittings $M_i$, and the suppressions
cannot be explained purely by SU(3)$_{\rm flavor}$-breaking effects.
The combined large $N_c$ (from either large $N_c$ limit) and
SU(3)$_{\rm flavor}$-breaking suppression factors give results much
more consistent with the data than do SU(3)$_{\rm flavor}$-breaking
factors alone.

Surprisingly, however, it is also clear that both the large
$N_c^{\mathrm{F}}$ and large $N_c^{\mathrm{AS}}$ expansions are
qualitatively similar in their match to the data: The spread of $A_i$
values in each case is only about 1.5 units.  Our results suggest that
baryon spectroscopy data are consistent with {\em two different large
$N_c$ limits}.  The analysis shows that large $N_c$ physics is
necessary to explain baryon mass splittings, but it does not
definitively show which large $N_c$ limit is closest to our $N_c=3$
world.

\section{A hybrid large $N_c$ limit}
\label{sec:CRlargeN}

In the last section we compared the phenomenology of two distinct
flavor-symmetric $1/N_c$ expansions extrapolated to $N_c=3$.  In this
section we discuss a hybrid large $N_c$ limit.  Specifically, suppose
one extrapolates to the large $N_c$ limit with one quark flavor in the
AS representation, but keeps the rest of the flavors in the F
representation.  This extrapolation has a number of theoretical
virtues connected to the large $N_c$ orientifold equivalence with
QCD$_{\rm Adj}$ described in the Introduction, and it has been
explored by Ref.~\cite{Armoni:2003gp, Kiritsis:1989ge,Frandsen:2005mb,Sannino:2007yp,
Armoni:2005wt,Armoni:2009zq,HoyosBadajoz:2009hb}.  We refer to this
large $N_c$ extrapolation, first suggested in 1979 by Corrigan and
Ramond (CR)~\cite{Corrigan:1979xf}, as the large $N_c^{\mathrm{CR}}$
limit.

Reference~\cite{Armoni:2003gp} observed that the F quarks decouple
from the dynamics of the glue and the AS quarks in the large
$N_c^{\mathrm{ CR}}$ limit, so that QCD in the CR large $N_c$ limit is
equivalent to $N_f = 1$ QCD$_{\rm AS}$ at large $N_c$.  Since the
latter theory is orientifold-equivalent to $\mathcal{N}=1$ SUSY for
large $N_c^{\rm Adj}$, one can use SUSY methods to make predictions
for the behavior of the theory in the large $N_c^{\mathrm{CR}}$ limit.
Thus, to the extent that large $N_c$ limit and the chiral limit of
this flavor-asymmetric theory are relatively close to the real world,
one obtains an {\it ab initio\/} method of predicting certain QCD
observables.  For instance, Ref.~\cite{Armoni:2005wt}, showed that the
$N_f=3$ chiral condensate $\langle \bar{q} q \rangle$ in the large
$N_c^{\mathrm{CR}}$ limit can be calculated using the orientifold
equivalence.

This theory has another virtue, noted long ago in
Ref.~\cite{Corrigan:1979xf}. The CR theory has states with the color
structure $Q^{[i,j]}q_i q_j$ at all $N_c$.  At $N_c=3$, using the AS
isomorphism described above, this state simply becomes the usual
three-quark baryon. If one assigns each fundamental quark $q_i, \,
q_j$ as well as $Q^{[i,j]}$ baryon number $1/3$ at general $N_c$, then
three-quark baryons $Q^{[i,j]}q_i q_j$ exist at any $N_c>1$.  Thus, in
contrast to Witten's construction in Eq.~(\ref{Witten}), CR baryon
masses do not grow with $N_c$, which is an attractive feature since
physical $N_c = 3$ baryon and meson masses are not significantly
different.

Moreover, CR baryons and mesons have recently been shown to exhibit a
very interesting property: Their Regge slopes
coincide~\cite{Armoni:2009zq}.  Since this feature approximately holds
in the physical $N_c=3$ hadronic spectrum, one might conclude that the
large $N_c^{\mathrm{CR}}$ limit is phenomenologically useful.

In the rest of this section we explore whether a $1/N_c$ expansion
around the large $N_c^{\mathrm{CR}}$ limit is actually
phenomenologically viable.  Clearly, one can construct a class of
theories at any $N_c$ with one flavor of AS quark and one or more F
quark flavors.  Such theories are certainly sensible at any $N_c$; at
$N_c=3$ they correspond to real QCD\@.  However, one may ask whether
the large $N_c^{\mathrm{CR}}$ world is ``close'' to the $N_c=3$ world.
In certain crucial aspects the answer is no.

For example, the approximate SU($2N_f$) $\times$ SU($2N_f$) chiral
symmetry is an essential feature of $N_c=3$ hadronic physics; let us
consider the fate of this symmetry in the large $N_c^{\mathrm{CR}}$
limit.  For simplicity we restrict to two massless flavors $u$ and
$d$, and without loss of generality we take $u$-flavor quarks to be AS
quarks, denoted by $u_{\mathrm{AS}}$, while $d$-flavor quarks
transform in the F representation, and are denoted by
$d_{\mathrm{F}}$.  For $N_c=3$, the exact SU(2)$_L \times$ SU(2)$_R$
chiral symmetry is spontaneously broken to SU(2)$_V$, yielding three
pseudoscalar Goldstone pions and a light $\eta^\prime$ associated with
the anomalous U(1) transformation $\exp (i\theta \gamma_5)$.  For any
finite $N_c \neq 3$ in the CR approach, no symmetry connects $u_{\rm
AS}$ and $d_{\rm F}$: The chiral symmetry and its associated Goldstone
bosons are present only in the $N_c = 3$ CR theory.

Indeed, one cannot construct color-singlet operators with the quantum
numbers of the pion for $N_c \ne 3$.  Consider an operator with the
quantum numbers of the $\pi^+$; for $N_c=3$ this is the color singlet
$i \epsilon^{j k l} \overline{u}_{j k} \gamma_5 d_l$.  For $N_c >
3$, $\epsilon^{j k l}$ is no longer an invariant tensor, and three
color indices cannot be combined into a color singlet.  Thus for any
finite $N_c > 3$, no Goldstone bosons occur.  As $N_c \rightarrow
\infty$, one light boson associated with the U(1) rotation $d_{\rm
F} \rightarrow \exp (i\theta \gamma_5) d_{\rm F}$, call it $\eta_d$,
emerges.  While this chiral current is anomalous, the anomaly is
suppressed by $1/N_c$.  However, the anomaly for the analogous current
for $u_{\rm AS}$ is unsuppressed at large $N_c$, since AS quark loops
are unsuppressed.  Thus the large $N_c^{\mathrm{CR}}$ world has one
$\eta_d$ boson, while the $N_c=3$ world has three $\pi$ Goldstone
bosons and an $\eta^\prime$.  This is an essential difference.

The problem with the large $N_c^{\mathrm{CR}}$ limit is not that chiral symmetry
is explicitly broken at $N_c>3$, but rather that the transformations
associated with the symmetry do not even \emph{exist\/} away from
$N_c=3$.  Moreover, isospin transformations also do not exist for $N_c
>3$.  Hadrons do not form isospin multiplets for $N_c>3$; for example,
while protons and neutrons both exist at $N_c=3$, only conventional
neutrons ($u_{\rm AS} d_{\rm F} d_{\rm F}$) exist for any $N_c >3$.
The large $N_c^{\mathrm{CR}}$ world is qualitatively {\em very\/}
different from the $N_c = 3$ world.  A phenomenologically viable
$1/N_c$ CR expansion seems unlikely for observables sensitive to
flavor physics.

The essential nature of the problem is the absence of a smooth limit.
Had the issue merely been that isospin and chiral symmetry were badly
broken at large $N_c^{\mathrm{CR}}$, then one might hope that the
symmetries at $N_c=3$ emerge when $1/N_c$ corrections are included to
sufficiently high order.  If, for example, isospin-violating
quantities such as the proton-neutron mass difference were substantial
at large $N_c$, but $1/N_c$ corrections reduced the difference in such
a way that their sum added to zero at $N_c=3$, one might be able to
construct a sensible $1/N_c$ expansion that is useful down to $N_c=3$.
However, since the proton simply does not exist for $N_c>3$, the
proton-neutron mass difference cannot be studied in a $1/N_c$
expansion.

Given these considerations, we conclude that the large $N_c^{\rm CR}$
limit, while exhibiting interesting Regge behavior, does not appear to
admit a usable $1/N_c$ expansion that connects smoothly to the $N_c=3$
world for a wide class of observables.

\section{Summary}
\label{sec:Conclusions}

In this paper, we have investigated the phenomenology of two large
$N_c$ limits of QCD: the large $N_c^{\mathrm{F}}$ and large
$N_c^{\mathrm{AS}}$ limits.  Using the emergent spin-flavor symmetry
of baryons in the large $N_c$ limit, we have compared the predictions
of both of these large $N_c$ limits for certain baryon mass relations
to experimental data.  We have found that these baryon mass relations
hold to the expected order for \emph{both} large $N_c$ limits.  Thus
at least for the class of observables we have considered, both large
$N_c$ limits appear to be phenomenologically viable.

\vskip 3ex
{\it Acknowledgements.}  We thank Adi Armoni for very helpful discussions and comments
on the manuscript.  We thank the organizers and
participants of the INT workshop ``New Frontiers in Large $N$ Gauge
Theories,'' where this work was initiated, and also the INT and the
University of Washington for their hospitality.  A.C.\ and T.D.C.\
acknowledge the support of the U.S.\ Dept.\ of Energy under Grant No.\
DE-FG02-93ER-40762.  R.F.L.\ acknowledges the support of the NSF under
Grant No.\ PHY-0757394.

\appendix
\section{Operator analysis and mass relations}

In this appendix we briefly sketch the arguments leading to the mass
relations given in Table~\ref{MassRelations}.

The phenomenological analysis of large $N_c$ baryons relies upon three
features: the large number of quarks in the baryon wave function
[$N_c$ in the $N_c^{\rm F}$ counting, $N_c(N_c-1)/2$ in the $N_c^{\rm
AS}$ counting] and their combinatorics~\cite{Witten:1979kh}, the
't~Hooft scaling~\cite{'tHooft:1973jz} $g_{\rm YM} \! \sim \!
N_c^{-1/2}$ of the QCD coupling constant, and the existence of a
ground-state multiplet completely symmetric under the combined
spin-flavor symmetry, which is the generalization of the old SU(6)
{\bf 56}-plet~\cite{Dashen:1993jt,Carone:1993dz,Luty:1993fu}.  Of
course, the baryon contains not only valence quark but gluon and sea
quark degrees of freedom as well.  However, the assumption that
physical baryons fill specific spin and flavor representations based
upon the quantum numbers of quarks implies that one may describe the
baryon wave function as comprised entirely of quark interpolating
fields that exhaust the baryon wave function~\cite{Buchmann:2000wf},
and it is these fields that one calls ``quarks'' in the large $N_c$
analysis.

Interactions among the quarks in the the baryon may be described in
terms of operators classified~\cite{Dashen:1994qi} by their
transformation properties under the spin-flavor symmetry (singlet,
adjoint under spin but not flavor, three-index symmetric tensor under
flavor, {\it etc.}) as well as the number $n$ of quarks that
participate in the interaction, which defines an {\it $n$-body
operator}.  The quark combinatorics combined with the 't~Hooft scaling
shows that $n$-body operators are generically suppressed by a factor
$1/N_c^n$ in the $N_c^{\rm F}$ counting. However, to obtain the true
suppression factor one must also take into account cases in which the
contributions from the quarks add coherently in the baryon matrix
elements, as well as eliminate operators whose matrix elements form a
linearly dependent set when evaluated on the ground-state baryon
multiplet~\cite{Dashen:1994qi}.  Once this is accomplished, one is
left with a linearly independent set of operators, each one with a
well-defined power counting in $1/N_c$, that carries precisely the
same dimension as the space of independent baryon observables. In
essence, the operators and observables form equivalent bases for the
baryons, and the operators form a hierarchy in powers of $1/N_c$ such
that their assembly forms an effective baryon Hamiltonian.

An explicit example clarifies the comments made above.  In either the
large $N_c^{\rm F}$ or $N_c^{\rm AS}$ limit, the color indices of the
quarks are combined to give a color wave function completely
antisymmetric under quark exchange, as in Eqs.~(\ref{Witten}) or
(\ref{psi}), respectively.  Since the baryons are fermions and
therefore completely antisymmetric under the exchange of all quark
indices, the spin-flavor-space part of the baryon wave function must
be completely symmetric under quark exchange; and since the
spatial wave functions of ground-state baryons are $L = 0$ and hence
symmetric, one requires complete symmetry of the spin-flavor wave
function.  This is the reasoning that, for $N_c = 3$, leads to the
SU(6) {\bf 56}-plet.

Operators that break the spin-flavor symmetry carry spin, flavor,
or both indices, and are labeled $J^i$, $T^a$, and $G^{ia}$,
respectively in
Refs.~\cite{Jenkins:1995td,Jenkins:2000mi,Dashen:1993jt}.  For
example,
\begin{equation}
T^a \equiv \sum_\alpha q^\dagger_\alpha \left( \openone \otimes
\frac{\lambda^a}{2} \right) q_\alpha \, ,
\end{equation}
where $\openone$ is the identity matrix in spin space, $\lambda^a$ is
the usual Gell-Mann matrix in flavor space, and $\alpha$ sums over all
the quarks in the baryon.  Note that color indices do not appear
explicitly in this expression, so the operator $T^a$ is well defined
in both large $N_c^{\rm F}$ and $N_c^{\rm AS}$.  All operators that
have nonvanishing matrix elements on baryon states may be formed from
polynomials of $J^i$, $T^a$, and $G^{ia}$ (with suitable contractions
of spin-flavor indices), and such a polynomial of $n$th order gives an
$n$-body operator.  Since the physical baryons have $N_c = 3$ quarks,
such polynomials beyond cubic order applied to these baryons give
matrix elements linearly dependent upon those of lower-order
operators, which means that the $1/N_c$ series for any finite given
value of $N_c$ terminates after providing a complete set of
independent operators.

An $n$-body operator requires an $n$-quark interaction, which in turn
implies $2n$ factors of $g_{\rm YM}$, for a suppression of $1/N_c^n$,
as discussed above.  Therefore, in the effective large $N_c^{\rm F}$
baryon Hamiltonian, the operator $T^a$ appears multiplied by an
explicit factor of $1/N_c$ compared to the spin-flavor symmetric
operator $\openone$ that has $O(N_c^1)$ matrix elements.  Consider,
however, the mass operator $T^8$; its matrix elements, which naively
merely count strange quarks, are actually given by
\begin{equation}
\langle T^8 \rangle = \frac{1}{2\sqrt{3}}(N_c - 3N_s) \, .
\end{equation}
Here we see a coherent $O(N_c^1)$ contribution that seems to upset the
large $N_c$ counting; however, note that it is the same for all
baryons and therefore simply provides an additional contribution to
the leading-order spin-flavor symmetric mass operator $\openone$.  The
operator $T^8$ also breaks SU(3)$_{\rm flavor}$ and therefore requires
an explicit prefactor of $\epsilon$.  When one repeats this analysis
for a complete set of linearly independent operators ({\it e.g.},
$\epsilon T^8$, $\epsilon^2 \{ T^8, T^8 \}/N_c$, {\it etc.}), one
finds that each operator contributes to a unique baryon mass
combination, which defines the combinations $M_i$ in
Table~\ref{MassRelations}.  For example, the combination $M_2$ is
associated with the operator $\epsilon T^8$ considered above.  The
analysis for $N_c^{\rm AS}$ proceeds exactly the same way, with the
exception that each $1/N_c$ in $N_c^{\rm F}$ is replaced by $1/N_c^2$.

Such an analysis is not unique to the $1/N_c$ expansion. All that is
required is a finite multiplet of states under some symmetry, and a
perturbative parameter that suppresses some of the independent
operators that act upon the multiplet.  For example, since the
fundamental operators distinguishing strangeness break
SU(3)$_{\mathrm{flavor}}$ symmetry by transforming as an {\bf 8},
flavor symmetry-breaking transforming as a {\bf 27} ($\subset {\bf 8}
\otimes {\bf 8}$) does not occur until second order in
SU(3)$_{\mathrm{flavor}}$ breaking, and this operator must be
associated with a doubly-suppressed SU(3)$_{\mathrm{flavor}}$-breaking
mass combination.  Indeed, when evaluated for the $N_c
\! = \! 3$ baryon octet, this mass combination turns out to be just
the one that defines the Gell-Mann--Okubo relation, $2N_0 -\Sigma_0
-3\Lambda +2\Xi_0$, using the notation of Table~\ref{MassRelations}.

This analysis, applied to the baryons in large $N_c^{\rm F}$ limit,
was first performed in Ref.~\cite{Jenkins:1995td}, and was later
improved as better measurements for baryon isospin splittings were
obtained~\cite{Jenkins:2000mi}.

\end{document}